\documentclass[twocolumn]{geophysics}  
\graphicspath{{Fig/}{Fig/}}
\usepackage{multirow}
\usepackage{array}
\usepackage{booktabs}
\usepackage{subfiles}
\usepackage{amsmath}
\usepackage{mathtools}
\usepackage{amsfonts}
\usepackage{amssymb}
\usepackage{listings}
\usepackage{indentfirst}  
\usepackage{ulem}
\usepackage{xcolor}
\DeclareMathOperator*{\argmin}{arg\,min}

\newcommand{\myvec}[1]{\text{\textbf{\lowercase{#1}}}}
\newcommand{\mymat}[1]{\text{\textbf{\uppercase{#1}}}}

\begin{document}

\onecolumn

\section{Copy Right Notice and Citation Information}
\noindent A revised version of this manuscript has been accepted to \textit{Geophysics} and is awaiting production checklist. The copyrights for the accepted manuscript belong strictly to the Society for Exploration Geophysicists (SEG). This document may strictly be used only for educational and other non-commercial purposes only. The full citation to the accepted manuscript will be made available once the DOI has been published. In the meanwhile, readers of the work are highly encouraged to cite this work using the following citation information:\newline\newline
\noindent \textbf{Full Text}: A. Mustafa, M. Alfarraj, and G. AlRegib, ``Joint Learning for Spatial Context-Based Seismic Inversion of Multiple Datasets for Improved Generalizability and Robustness," Geophysics, submitted on Jun. 19 2020.\newline\newline
\noindent \textbf{Bibtex}: \begin{lstlisting}
@misc{jointlearninginversion,
      title={Joint Learning for Spatial Context-based Seismic Inversion of Multiple 
      Datasets for Improved Generalizability and Robustness}, 
      author={Ahmad Mustafa and Motaz Alfarraj and Ghassan AlRegib},
      year={2021},
      eprint={},
      archivePrefix={arXiv},
      primaryClass={eess.IV}
     } 
\end{lstlisting}

\section{Author Information}
\begin{enumerate}
    \item[1.]  Ahmad Mustafa, Center for Energy and Geo Processing (CeGP), School of Electrical and Computer Engineering, Georgia Institute of Technology, Atlanta, GA, 30332, amustafa9@gatech.edu
    \item[2.] Motaz Alfarraj, King Fahd University of Petroleum and Minerals, Dhahran, Saudi Arabia, motaz@kfupm.edu.sa
    \item[3.] Ghassan AlRegib, Center for Energy and Geo Processing (CeGP), School of Electrical and Computer Engineering, Georgia Institute of Technology, Atlanta, GA, 30332, alregib@gatech.edu 
\end{enumerate}

\section{Code and Data Availability}
\noindent \textbf{Github Link}: https://github.com/amustafa9/Joint-learning-with-spatial-context-for-inversion \newline\newline
\noindent For issues with the code and/or accessing the data used for the results, please contact the primary author, Ahmad Mustafa.

\newpage

\twocolumn

\title{Joint Learning for Spatial Context-based Seismic Inversion of Multiple Datasets for Improved Generalizability and Robustness}

\renewcommand{\thefootnote}{\fnsymbol{footnote}} 

\address{
\footnotemark[1]Center for Energy and Geo Processing (CeGP), Omni Lab for Intelligent Visual Engineering and Science (OLIVES), School of Electrical and Computer Engineering, Georgia Institute of Technology, Atlanta, GA.
\footnotemark[2]King Fahd University of Petroleum and Minerals, Dhahran, Saudi Arabia.
}
\author{Ahmad Mustafa\footnotemark[1], Motaz Alfarraj\footnotemark[2], and Ghassan AlRegib\footnotemark[1]}

\footer{Example}
\lefthead{Mustafa \& AlRegib}
\righthead{Robust Learning of Seismic Inversion}

\begin{abstract}
\label{sec:abstract}
Seismic inversion plays a very useful role in  detailed stratigraphic interpretation of seismic data. Seismic inversion enables estimation of rock properties over the complete seismic section. Traditional and machine learning-based seismic inversion workflows are limited to inverting each seismic trace independently of other traces to estimate impedance profiles, leading to lateral discontinuities in the presence of noise and large geological variations in the seismic data. In addition, machine learning-based approaches suffer the problem of overfitting if there is a small number of wells on which the model is trained. We propose a two-pronged strategy to overcome these problems. We present a Temporal Convolutional Network that models seismic traces temporally. We further inject spatial context for each trace into its estimations of the impedance profile. To counter the problem of limited labeled data, we also present a joint learning scheme whereby multiple datasets are simultaneously used for training, sharing beneficial information among each other. This results in the improvement in generalization performance on all datasets. We present a case study of acoustic impedance inversion using the open source SEAM and Marmousi 2 datasets. Our evaluations show that our proposed approach is able to perform robust and laterally consistent estimations of impedance in the presence of noisy seismic data and limited labels. We compare and contrast our approach to other learning-based seismic inversion methodologies in the literature. On SEAM, we are able to obtain an average MSE of 0.0966, the lowest among all other methodologies.
\end{abstract}

\section{Introduction}
\label{sec:intro}
Seismic inversion refers to the process of estimating reservoir properties from seismic reflection data. While the rock properties can be measured directly at the wells, they have to be estimated away from well locations using seismic data. Seismic inversion plays an important role in seismic interpretation. While the positions of major reflectors can generally be picked out in migrated, post-stack seismic sections, it is hard to identify the layer lithologies without knowing the rock properties in those regions. Detailed stratigraphic interpretation of a seismic section usually involves its inversion to an acoustic impedance section \citep{Gluck1997}. Acoustic impedance, being a reservoir parameter, can be related more easily to rock property measurements obtained through well logging. It allows delineation of major geologic changes in the subsurface since rocks of a similar type would have similar rock property values \citep{Lindseth1979}. Detailed analysis of the inverted parameters also enables the identification of the different rock types making up the subsurface. The resultant building of more accurate and reliable subsurface models helps with oil and gas exploration and production purposes. 

Seismic inversion operates on migrated seismic data and can be performed either on post-stack or pre-stack seismic sections. Some of the steps involved in seismic inversion are 1) pre-processing of seismic data to remove the effects of multiples, transmission losses, wave dispersion phenomena etc., 2) well-seismic ties and extraction of the wavelet, and 3) running the inversion algorithm itself. Model-based seismic inversion works by starting with a smooth model of the subsurface. A synthetic seismic response is obtained from this model in a process called forward modelling. The synthetic seismic response is compared to the actual seismic, and the error is used to update the model parameters. Multiple iterations of this process are performed until the synthetic seismic matches the actual seismic response to an acceptable degree of accuracy. This optimization procedure can be mathematically expressed as follows: 

\begin{equation}
    \hat{m} = \argmin_{m} \quad \mathcal{L}(f(m), d) + \lambda\mathcal{C}(m),
    \label{eq:1}
\end{equation}
where $f( m)$ represents the synthetic seismic generated by forward modelling on the model parameters, $m$. $\mathcal{L}(f(m), d)$ represents some distance measure between the synthetic seismic and the actual seismic $d$. $C(m)$ represents a regularization term imposed upon the problem to deal with the non-uniqueness of the solutions, while $\lambda$ is the weight given to this term. $\hat{m}$ refers to the optimal solution found for the optimization problem. A comprehensive survey of the various seismic inversion methods is given in \cite{Veeken2004SeismicIM}.

Deep learning, a subset of machine learning, has in the recent past led to ground breaking advancements in the field of image classification  \citep{Krizhevsky2017}, object detection \citep{objdetect}, image segmentation \citep{segmentation}, image and video captioning \citep{captioning}, speech recognition \citep{speech}, and machine translation \citep{DBLP:conf/emnlp/ChoMGBBSB14}. The success of deep learning in computer vision and natural language processing domains has of late inspired geophysicists to replicate these successes in the field of seismic interpretation. Machine learning has been used to solve problems in salt body delineation \citep{haibinSaltbodyDetection, AsjadSaltDetection, AmirSaltDetection,Wang2015}, fault detection \citep{haibinFaultDetection, HaibinFaultDetection2,Di2019Fault} , facies classification \citep{YazeedFaciesClassification, YazeedFaciesWeakClassification},  seismic attribute analysis \citep{Long2018TextureAnalysis, Di20183DcurvatureAnalysis, Alfarraj2018Multiresolutionanalysis}, and structural similarity based seismic image retrieval and segmentation \citep{YazeedStructurelabelPrediction}.

Seismic inversion has also been attempted in the past using various machine learning-based approaches. A non-linear mapping, $\mathcal{F}$, characterized by a set of parameters, $\Theta \in \mathbb{R}^{n}$ is learnt from a training dataset ($\mathcal{D}$) consisting of well-logs as labels ($Y$) and the seismic traces corresponding to those well-logs as features ($X$). Once the machine learning model has been trained, it is used to invert the seismic traces at all the non-well positions to obtain a rock property volume. This is expressed more succinctly in the equation below:
\begin{equation}
    \mathcal{F}_{\Theta}:X\xrightarrow{}Y.
    \label{eq:ml_based_inverrsion}
\end{equation}

Artificial feed-forward neural networks (ANNs) feature prominently in the literature involving such approaches \citep{Banchs2002, Hampson2001, Liu1998, Rth1994}.
Artificial neural networks are able to learn highly complex, non-linear mappings given a sufficiently large set of labeled training samples. However, ANNs are prone to overfitting in the presence of limited labeled data. More recently, Convolutional Neural Networks (CNNs) were used for rock property estimation from seismic data \citep{BiswasPhysicsGuidedCNN, DasCNNInversion}. CNNs are a type of neural network that learn complex mappings between the input and output domains by sliding a set of trainable convolutional kernels over the input features. This sharing of weights in a CNN results in it having fewer parameters than an ANN of the same depth, making it less vulnerable to overfitting. Moreover, it allows the construction of deeper networks to learn richer representations. 

Around the same time, \cite{motazRNN1} showed that by capturing the temporal relationships in seismic traces via a hidden state vector, Recurrent Neural Networks (RNNs) were able to efficiently estimate rock properties without actually requiring large amounts of training data, as is common with other non-sequence modelling based neural network architectures. Shortly afterwards, \cite{MustafaTCN} introduced another kind of sequence modelling neural network based on Temporal Convolutional Network (TCN) for estimation of acoustic impedance (AI) from seismic data. By explicitly modeling seismic data and the well logs as time-series data, such sequence modeling-based approaches are able to produce accurate estimates of well pseudologs from limited training data. 

To utilize the large amounts of unlabeled seismic data in learning-based inversion schemes and also to constrain the neural network to produce more regularized estimates of rock property pseudologs,  \citep{motazSemiSupervisedAcoustic, motazSemiSupervisedElastic} introduced a network architecture based on both RNNs and CNNs. They injected physical constraints into the network, producing an accurate estimate of the Elastic Impedance section from a small-sized, labeled training dataset. 

A drawback faced by classical and deep learning-based seismic inversion workflows is that each seismic trace is inverted independently of other traces. However, in a seismic image of the subsurface, neighbouring traces are highly correlated. A property estimation approach working on a trace-by-trace basis, is not able to take this information into account. This can lead to lateral discontinuities in the inverted property volumes, especially in the presence of noise. However, naively extending the neural network architecture to include neighboring seismic traces as features without paying regard to the ordering of the traces in the seismic image might not help the network; it might even make matters worse since now the network has more parameters to learn. 

The problem is compounded by the fact that any given seismic survey is likely to contain only a few wells, given the high cost of drilling. Fitting a machine learning model on surveys with limited training data makes it vulnerable to overfitting. This bottleneck caused by limited labeled data may be circumvented by using well-logs from other seismic surveys. Transfer learning is a popular machine learning paradigm utilized to improve generalization performance in problems where there might be too few labeled data available for a given task. Useful feature representations learned on a large, labeled source dataset are transferred to a target dataset with fewer labeled training samples. Deep neural networks are known to learn general, domain-invariant features in their earlier layers, followed by more specialized, dataset- and task-specific features in the later parts of the network \citep{transferlearningYosinski2014}. The use of deep neural networks pretrained on large image datasets as feature extractors for other machine learning models and tasks has been well investigated \citep{MohanNeuralNetworkFeatureExtraction,notley2018NeuralNetworkFeatureExtractors}. 

A common transfer learning methodology for deep neural networks is to transfer the weights of the first few layers of a network trained on a labeled source dataset to the corresponding layers of a second network with an identical architecture \citep{transferlearningYosinski2014}. With the layers containing transferred weights frozen in place, the remaining layers of the network are randomly initialized and trained in a supervised fashion on the few labeled training samples in the target dataset. Alternatively, the weights in the pretrained layers may also be allowed to change with a smaller learning rate compared to the later layers in the network. This pretraining strategy has the drawback that it requires trial and error and/or domain expert knowledge to determine the optimal number of pretrained layers for a given neural network architecture trained for a certain task. The option to either keep the pretrained weights frozen or let them change with a certain learning rate is another hyperparameter that has to controlled by the user. Moreover, as the discrepancy between the source and target datasets increases, it may become counterproductive to use pretrained weights from the source dataset. In the specific context of learning-based seismic inversion, finetuning network weights after pretraining on a source well log dataset leads to suboptimal network performance on the original source task.

We propose a two-pronged strategy to combat the problems faced by deep learning-based seismic inversion methods described above. We present a novel network architecture that models seismic data both \emph{temporally} and \emph{spatially}. Our network is derived from the CNN-based sequence modelling architecture introduced in \cite{Bai} and further adapted to seismic inversion in \cite{MustafaTCN}, called a Temporal Convolutional Network (TCN). We extend and build upon this to introduce a two dimensional TCN-based architecture that is able to not only learn temporal relationships within each seismic trace, but also inject spatial context from neighboring traces into the network estimations. It does this by processing each data instance as a rectangular patch of seismic image centered at the well position, rather than just the single seismic trace at the well. By processing seismic data in this way, we are able to preserve and utilize the spatial structure of seismic images and still be able to model temporal relationships in seismic traces. As we show later, this leads to better lateral continuity in the estimated rock property sections in the presence of noisy seismic data.

To overcome the problem of limited training data and the limitations of conventional pretraining-based transfer learning approach commonly used in such scenarios, we propose a bi-directional transfer learning strategy to simultaneously learn on multiple datasets, leading to an improvement in generalization performance on all datasets. Our proposed approach is able to dynamically share mutually useful knowledge between tasks without imposing any hard constraints on the number of shared, pretraining layers on the network. In the presence of noisy seismic data and a limited number of well logs, we demonstrate that our proposed approach is able to perform more robust estimations of rock property pseudologs compared to approaches that only model seismic traces as one-dimensional sequences. We achieve a higher generalization performance on all datasets compared to if we used only training samples from any one dataset. We also report our approach to be user friendly in terms of not requiring trial and error to best determine pretraining hyperparameters, as in conventional transfer learning with deep neural networks.

In short, our contributions in this work is two-fold:
\begin{itemize}
    \item We present a neural network capable of simultaneously modeling the spatial and temporal relationships in seismic traces to produce robust estimates of rock property sections. The estimated sections show greater continuity and improved resolution over approaches that only model seismic traces temporally.
    \item We propose a transfer learning scheme where we learn on labeled data from multiple surveys, improving the generalization performance on all surveys. Compared to pretraining on a source dataset, our approach has the advantage of dynamic bi-directional knowledge sharing with minimal user input to lead to optimal performance on all datasets. 
\end{itemize}

\section{From ANNs to Sequence Models}
\label{sec:sequence_modelling}
Feedforward Neural Networks are capable of learning complex, non-linear relationships between a set of input and output domains, if provided with sufficient labeled training data. They consist of a layer of input nodes, followed by one or several layers of hidden nodes, followed lastly by a layer of output nodes. Each layer is connected to the next via dense feed-forward connections. The output activations from each layer are a weighted sum of activations in all the nodes of the previous layer, followed by the application of a non-linear function. This is shown in Equation \ref{eq:feed_forward} below:

\begin{equation}
    \myvec{o}^{l} = g(\mymat{W}^{l}\myvec{o}^{l-1} + \myvec{b}^{l}),
    \label{eq:feed_forward}
\end{equation}

\noindent where $\myvec{o}^{l}$ denotes the output activations of layer $l$, $\myvec{o}^{l-1}$ the vector of output activations in layer $l-1$, $\mymat{W}^{l}$ the matrix of weights connecting the two layers, and $\myvec{b}^{l}$ the vector of biases in layer $l$. $g(.)$ is a non-linear function, a popular choice for which is the Rectified Linear Unit (ReLU), shown in Equation \ref{eq:relu} below:

\begin{equation}
    g(x) = \text{max}(0, x).
    \label{eq:relu}
\end{equation}
 
Since every layer in a feedforward Neural Network is characterized by a dense matrix of weights, these neural networks can quickly become over-parameterized if built too deep. This makes them vulnerable to overfitting. Moreover, they do not explicitly encode the temporal and/or the spatial relationships in input features if the input happens to be structured, e.g., images, speech, and other time-series data.   

Recurrent Neural Networks (RNNs) \citep{Sherstinsky_2020}, are a class of neural networks especially designed to process sequence data. They explicitly model their input, $\{\myvec{x}_{t}\}_{t=1}^{T}$ to be a sequence consisting of disparate points in time, where $\myvec{x}_{t} \in \mathbb{R}^{d}$. At each time step, $t$, a hidden state vector, $\myvec{h}_{t}$ is computed, which is then passed on to the next time step. This is shown in Equation \ref{eq:rnn_hidden_state} below:

\begin{equation}
    \myvec{h}_{t} = g(\mymat{W}_{xh}\myvec{x}_{t} + \mymat{W}_{hh}\myvec{h}_{t-1} + \myvec{b}_{h}),
    \label{eq:rnn_hidden_state}
\end{equation}

where $\myvec{h}_{t-1}$ is the hidden state vector from the previous time step, $\myvec{x}_{t}$ is the input feature vector at time step $t$, and $\mymat{W}_{xh}$, $\mymat{W}_{hh}$, and $\myvec{b}_{hh}$ are matrices and the bias vector respectively associated with the feedforward computations. $g(.)$ denotes the application of a non-linear function, as described before.

The hidden state is then used to compute the output, $y_{t}$ at time step $t$ via feedforward connections characterized by $\mymat{W}_{hy}$ and $\myvec{b}_{y}$:

\begin{equation}
    \myvec{y}_{t} = g(\mymat{W}_{hy}\myvec{h}_{t} + \myvec{b}_{y}). 
    \label{eq:rnn_output}
\end{equation}

Moreover, in an RNN, the weights characterized by  $\mymat{W}_{xh}$, $\mymat{W}_{hh}$, $\myvec{b}_{hh}$, $\mymat{W}_{hy}$, and $\myvec{b}_{y}$ are shared at each time step. This results in a highly specialized network architecture that can not only scale to long sequences, but also use past history of inputs to make more informed output estimations at each time step.

Another class of neural networks that is able to model its input as a temporal sequence is a 1-D Convolutional Neural Network . It consists of a set of trainable convolutional kernels that slide over an input sequence to generate an output sequence. By employing parameter sharing concepts similar to those in an RNN, each activation of the output at a certain time step is a function of a small number of elements in the input sequence at neighboring time steps. By stacking together multiple convolutional layers, one can expand the receptive field of the network, enabling it to look at a longer history of the input to produce an activation at a certain time step of the output. For a one-layer CNN, the relationship between input and output is given by:

\begin{equation}
     y[t] = g(\sum_{m} x[m]w[t-m]),
     \label{eq:cnn}
\end{equation}

where $y[t]$ represents the activation of the output sequence $y$, at time step $t$, $x$ the input sequence, and $w$ the convolutional kernel. $g(.)$ as before, represents the application of a non-linear function.  

\section{Temporal Convolutional Network}
\label{sec:1D_TCN}
The authors in~\cite{MustafaTCN} propose a neural network for seismic inversion based on the Temporal Convolutional Network Architecture (TCN) described in \cite{Bai}. The TCN possesses a number of specialized features that enable it to efficiently model its input as a sequence.

\subsection{1-D Temporal Convolutional Network}
\label{subsec:1D_TCN}
The Temporal Convolutional Network (TCN) architecture consists of a series of temporal blocks. Each temporal block receives a 1-D input of a certain number of channels. The output of the temporal block, also a 1-D signal having a certain number of channels, is passed to the next temporal block. Below, we describe the salient features underpinning the TCN architecture.

\subsubsection{Temporal Block}
\label{subsubsec:temporal_block}
\begin{figure}
    \centering
    \includegraphics[width=0.5\columnwidth]{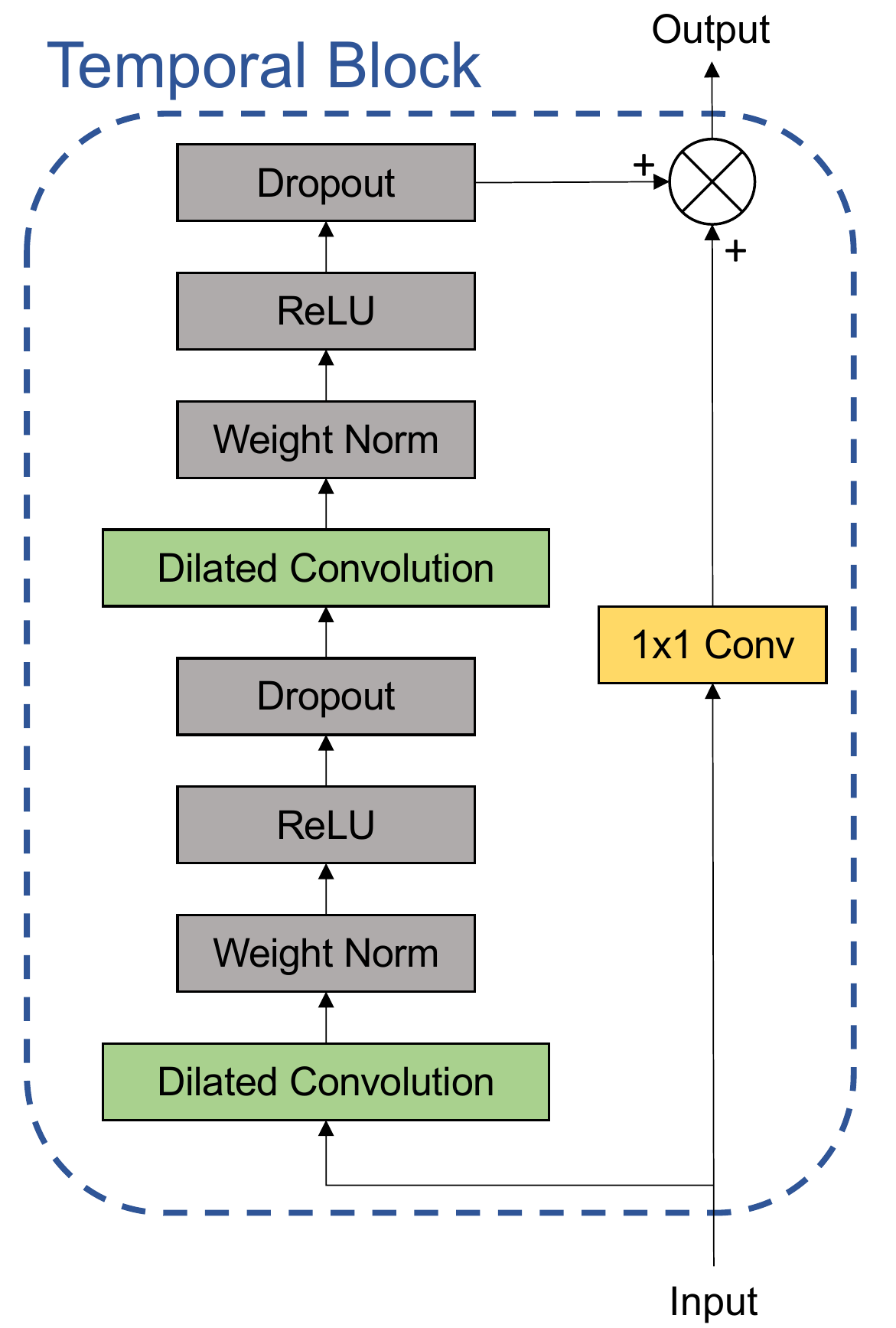}
    \caption{The basic structure of a Temporal Block. It consists of a mixture of convolutional, rectified non-linearity (ReLU), weight normalization, and dropout layers.}
    \label{fig:TemporalBlock}
\end{figure}

Figure \ref{fig:TemporalBlock} shows the structural composition of a Temporal Block. The Temporal Block is the basic building block of a Temporal Convolutional Network (TCN). Each Temporal Block has the same basic structure; it consists of convolutional layers interspersed with weight normalization, Dropout, and Rectified Non-linearity (ReLU) layers. The convolution layers extract features from the input by convolving it with a set of 1-D kernels. The input is padded before being processed by the kernels so that the output and input lengths stay the same. ReLU layers introduce a non-linearity into the network activations so that more powerful representations may be learnt. Dropout layers randomly zero out network activations to prevent the network from overfitting. Weight normalization layers, as described in \citep{weight_normalization}, reparameterize the weight vectors to have decoupled length and direction components. This helps to improve the convergence of the optimization during network training. There is also a skip connection from the input of the temporal block to the output. Adding skip connections in this way, as shown by \citep{resnet}, also helps stabilize network training and achieve better convergence. 

\subsubsection{Dilated Convolutions}
\label{subsubsec:dilated_convolutions}
Another major feature TCNs possess is dilated convolutions. As described in \cite{MultiScaleCA}, successive pooling operations cause a loss of input resolution in a CNN. This results in performance degradation for problems requiring prediction of dense outputs, such as semantic segmentation. However, pooling helps to aggregate global contextual features also needed for accurate output estimation. Dilated convolutions get around this problem by processing the input with sparse kernels, with the sparsity increasing exponentially along the network depth. This helps to capture global features while simultaneously preventing the loss of resolution in the network activations. In addition to this, it helps overcome overfitting by not requiring the network to be very deep to fully capture the global context in the input. Dilated convolutions effectively expand the receptive field of the network while preserving input resolution. The TCN utilizes 1-D dilated convolutions to better capture long term dependencies in the input seismic trace and produce more accurate property estimations.     

\subsection{2-D Temporal Convolutional Network}
\label{subsec:2D-TCN}
In this work, we adapted the architecture described in the previous subsection to process a 2-D image of seismic data centered at the well position rather than just the 1-D seismic trace there. The main features of the architecture are described below.

\subsubsection{Spatiotemporal Modeling of Seismic Data}
\label{subsubsec:spatiotemporal_model}
\begin{figure*}
    \centering
    \includegraphics[width=\textwidth]{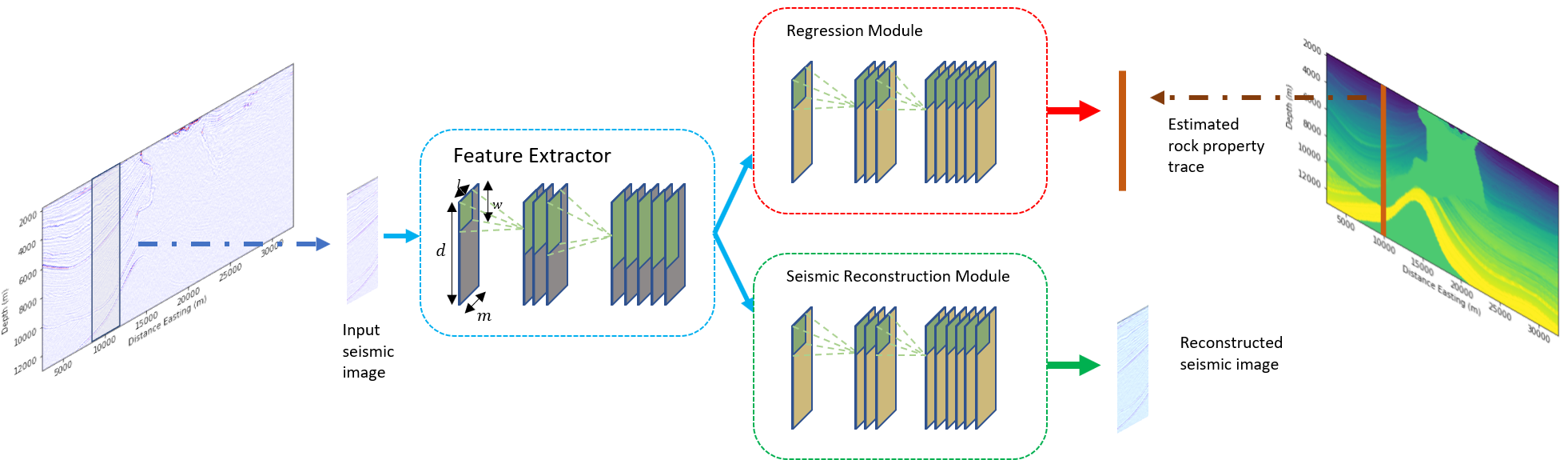}
    \caption{The network architecture. A seismic image of dimensions $d\times m$ is processed by the feature extractor block that consists of multiple 2-D temporal blocks. The 2-D kernels stay fixed in length $l$ but increase exponentially in width $w$. The features extracted by this block are output simultaneously to two different shallow 2-D CNNs to output estimated property trace and reconstructed input seismic image respectively.}
    \label{fig:network}
\end{figure*}

\begin{figure}
    \centering
    \includegraphics[width=\columnwidth]{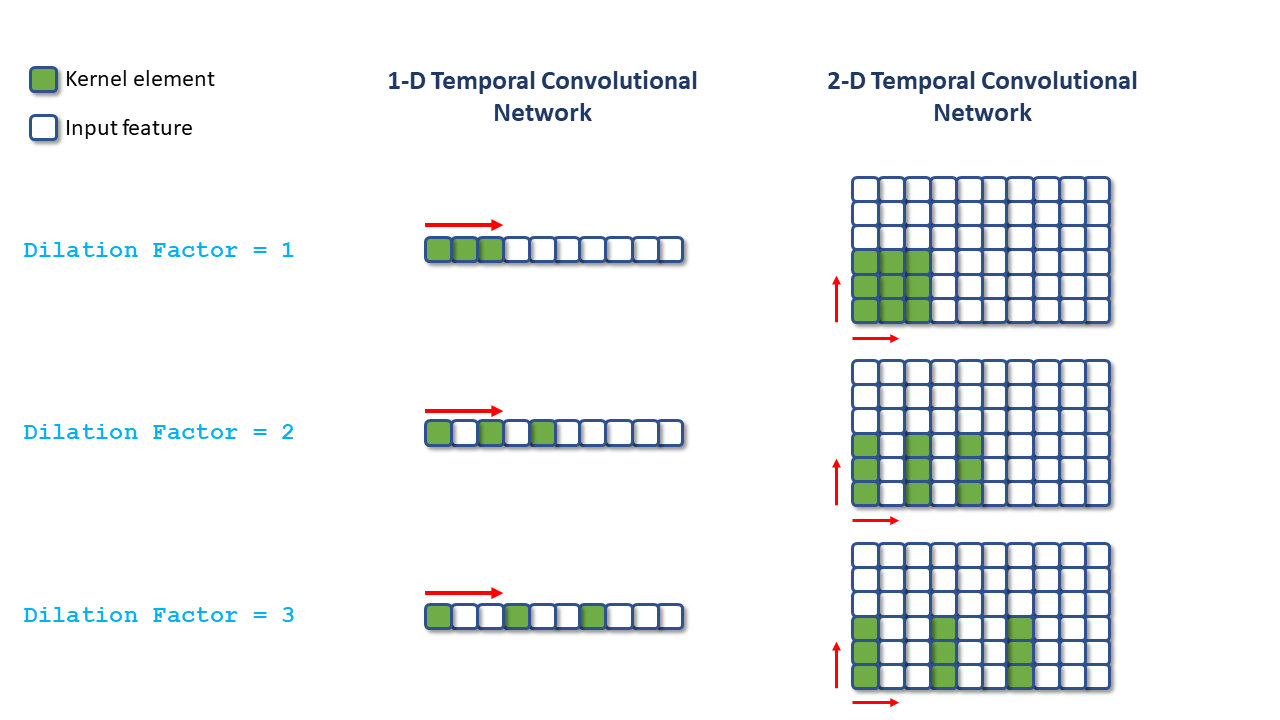}
    \caption{2-D convolutions in the proposed architecture explained and contrasted with their 1-D counterpart.}
    \label{fig:2_d_convolutions}
\end{figure}

As described in \cite{MustafaTCN}, the main body of the network, the feature extractor is made up of multiple temporal blocks, except that in this work, these are based on 2-D rather than 1-D convolutions. The basic structure of the temporal block stays the same as shown in Figure \ref{fig:TemporalBlock}. The network is input a rectangular patch of seismic data centered at the well position. The feature extractor block in the network processes this seismic image using 2-D kernels in each layer. The kernel dimensions increase exponentially in depth because of dilation but stay constant in the spatial direction. The exponentially increasing dilation factor results in temporal modelling of seismic traces for better estimation of the corresponding well log properties. The kernel being 2-dimensional helps to incorporate the local spatial context into the network estimations. This is demonstrated in Figure \ref{fig:2_d_convolutions}. The number of channels is increased after each temporal block, helping to learn better features. The extraction of the seismic images corresponding to each rock property trace and their subsequent processing via the aforementioned 2-D kernels is shown in Figure \ref{fig:network}. 

As we show later, the injection of spatial context considerably improves the quality of our network estimations, as may be observed in Figure \ref{fig:seam_plots}(b) and Figure \ref{fig:seam_plots}(h) respectively. 

\subsubsection{Regression and Reconstruction Modules}
\label{subsubsec:reg_and_recons_modules}
The output activations of the feature extractor 
are shared between the regression and reconstruction blocks, both of whom consist of three layers each of 2-D convolutions. The regression module processes these activations to produce the final property log at the well position. On the other hand, the reconstruction module uses them to reconstruct the input seismic image. This is an example of multi-task learning where the network is learning two tasks at once: property log estimation and seismic input reconstruction. By sharing representations between the two tasks, especially if they are related to each other, we bias the network to learn more generalizable features. For an overview on multitask learning, refer to \cite{ruder2017overview}. Multi-task learning potentially results in the network to perform better on all tasks than if trained to do each separately. The regression and reconstruction modules can be observed in Figure \ref{fig:network}.

\section{Joint Learning with Weight Sharing}
\label{sec:weight_sharing}
The 2-D TCN architecture just described leads to the network incorporating local spatial context in a migrated seismic image in its estimations of the rock property pseudologs. This is a means of extracting more information in a given seismic survey to help with the inversion. However, if there are only a few wells in the survey, injecting spatial context will only take the network so far. This is where knowledge of well logs in other surveys can be used to improve generalizability of the network. 

It is a common understanding among deep learning experts that deep neural networks learn a hierarchy of feature representations, from simple general features in the earlier layers to advanced, more specific ones in the deeper layers. Intuitively, different datasets should share to some extent the features in the earlier layers, before expressing specialized feature representations in the later layers. A common transfer learning methodology in the context of deep neural networks utilizes this idea by pre-training a network on a large source dataset before finetuning it on the usually much smaller target dataset. During the finetuning stage, the weights in the earlier, pre-trained layers are either kept fixed or only allowed to change at a small pace. In contrast, the deeper layers are allowed to adapt to the training examples from the target dataset at a higher learning rate. However, deciding the optimal number of pre-trained layers for a particular problem setup requires domain knowledge and/or a trial and error approach. Moreover, finetuning a network on the target dataset usually results in the degradation of network performance on the source. 

Another kind of transfer strategy learning may involve aggregating training examples from different datasets to form a new, bigger dataset. A single neural network may then be trained on the new dataset to incorporate knowledge from all the different components making it up. A drawback with such a strategy is that it is hard to control network behavior. For example, if one of the components only makes up a small fraction of the dataset, the network may simply choose to overfit on these training samples. As long as it performs well on training samples from the larger component dataset, it would still be able to achieve a lower loss. The network may or may not achieve better generalization; it is hard to control and predict. 

\begin{figure}
    \centering
    \includegraphics[width=0.95\columnwidth]{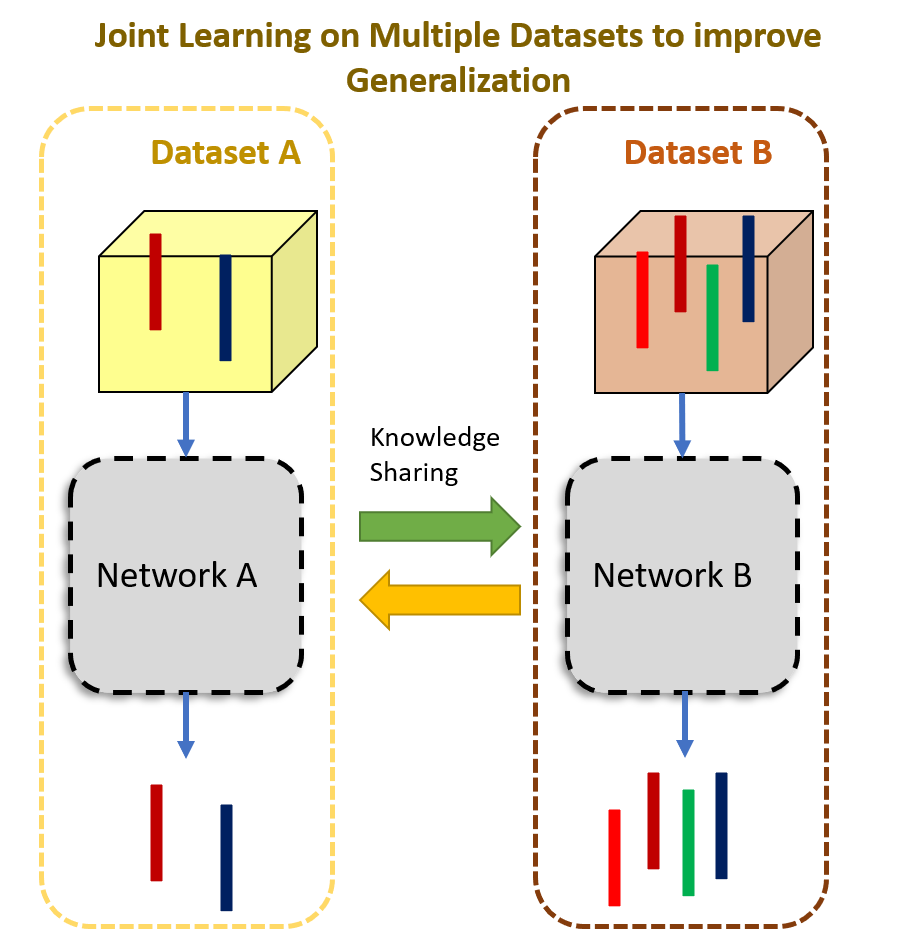}
    \caption{Learning simultaneously on two datasets. Each only acquires beneficial information from the other.}
    \label{fig:knowledge_share}
\end{figure}

In contrast to the two approaches for transfer learning described above, we propose and describe a knowledge sharing scheme whereby the generalization performance for seismic inversion is improved on all the different surveys involved in the experiment. This is done by letting the different tasks dynamically learn from each other only when it is useful while simultaneously giving them the room to optimize on their own respective datasets. It comes with the added advantage that the user does not have to pre-define/hard-encode the number of pre-training layers. Moreover, since each task has the flexibility to optimize on its dataset while learning from other tasks, we ensure that the generalization performance will not drop below the case if we trained each task on its dataset alone. This idea is illustrated in Figure \ref{fig:knowledge_share}.

\begin{figure}
    \centering
    \includegraphics[width=0.95\columnwidth]{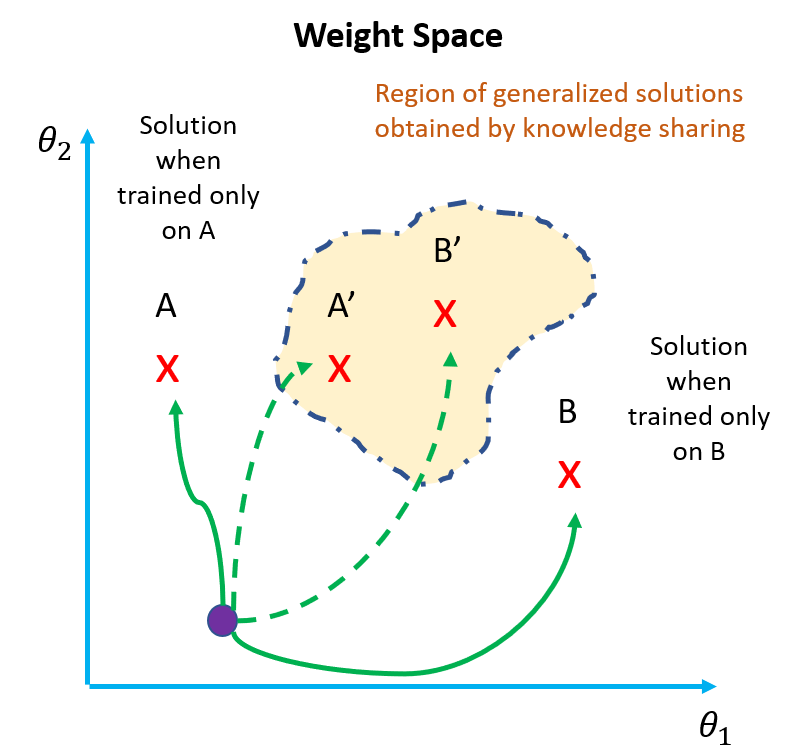}
    \caption{Exploration of the parameter space by the two networks, with and without the proposed knowledge sharing scheme.}
    \label{fig:weight_share}
\end{figure}

 We take two identical copies of the 2-D TCN architecture described previously and shown in Figure \ref{fig:network}, and train them simultaneously on the two datasets. In addition to optimizing the losses between the network estimations of well log property and the reconstructed seismic input with their respective ground-truths, we also minimize the squared L2 norm between the weights in all the corresponding layers in the two networks. By doing this, we effectively bias the networks to search the parameter space for a common region of solutions where the architecture will generalize better to inputs sampled from different distributions. However, by not constraining the weights to be exactly the same, each copy of the architecture is also free to find the optimal set of weights for its respective dataset in the vicinity of this solution space. Moreover, in the situation where the two datasets are very different from each other and learning on one will not help the other, the networks can choose to not learn from each other at all. The process is illustrated in Figure ~\ref{fig:weight_share}. Consider the two networks to be represented by $\mathcal{F}$ and $\mathcal{G}$ respectively. Both $\mathcal{F}$ and $\mathcal{G}$ consist of trainable weights organized into a set of $L$ convolutional layers. Consider $\theta_{A}^{l}$ to be the weight tensor in the $l$-th layer in network $A$, where $l\in [0, L-1]$. Then both $\mathcal{F}$ and $\mathcal{G}$ can be represented as follows:

\begin{equation}
    \mathcal{F} = [\theta_{\mathcal{F}}^{0}, \theta_{\mathcal{F}}^{1},\cdots, \theta_{\mathcal{F}}^{L-1}].
    \label{eq:network1}
\end{equation}

\begin{equation}
    \mathcal{G} = [\theta_{\mathcal{G}}^{0}, \theta_{\mathcal{G}}^{1},\cdots, \theta_{\mathcal{G}}^{L-1}]. 
    \label{eq:network2}
\end{equation}

The Weight Mismatch Loss is then defined as:
\begin{equation}
    l_{WML} = \sum_{l=0}^{L-1} \|\theta_{\mathcal{F}}^{l} - \theta_{\mathcal{G}}^{l}\|^{2}.
    \label{eq:weight_mismatch}
\end{equation}

By having the networks share weights , we force the two networks to learn from each other while they are optimizing the losses on their respective outputs. This rests on the assumption that there is a region of solutions in the shared parameter space that is able to generalize better to seismic data from multiple distributions. We force the networks to look for a solution in this space during training while simultaneously giving them room to optimize on their respective datasets. The process is illustrated in Figure \ref{fig:weight_share}. In the simplified 2-D parameter space shown in the figure, $\theta_{1}$ and $\theta_{2}$ may be considered to be the only weight parameters of the machine learning model. A and B are solutions the networks would have looked for if trained independently. Training jointly, they are forced to look for a solution space of greater generalization in proximity to each other.

\section{Network Training}
\label{sec:network_training}
Consider $D_{1} = \{X_{1}, Y_{1}\}$ and $D_{2} = \{X_{2}, Y_{2}\}$ to represent our two datasets, where the subscript refers to the dataset. $X = \{x^{1}, ..., x^{N}| x^{i} \in \mathbb{R}^{d \times m}\}$ represents the collection of $N$ seismic images in a dataset, where each $x^{i}$ is a $d\times m$ dimensional image. $d$ refers to the depth of the image while $m$ is the width. $Y = \{y^{1}, ..., y^{N}|y^{i}\in\mathbb{R}^{d}\}$ refers to collection of well log properties corresponding to each $x^{i} \in X$, where each $y^{i}$ is a $d$-dimensional rock property trace.
A batch of seismic images from each dataset is processed by its respective network to get the estimated well properties, $\hat{y}^{i}$ as well as the reconstructed seismic images, $\hat{x}^{i}$ as shown below:

\begin{equation}
    \hat{y}_{1}^{i}, \hat{x}_{1}^{i} = \mathcal{F}_{\Theta}(x_{1}^{i}). 
\label{eq:forward_modeling_eq_1}
\end{equation}

\begin{equation}
    \hat{y}_{2}^{i}, \hat{x}_{2}^{i} = \mathcal{G}_{\Theta}(x_{2}^{i}). 
\label{eq:forward_modeling_eq_2}
\end{equation}

The regression and reconstruction losses are then defined as:
\begin{equation}
    l_{reg} = \frac{1}{N_{1}}\sum_{i=1}^{N_{1}}\|\hat{y_{1}}^{i} - y_{1}^{i}\|_{2}^{2} + \frac{1}{N_{2}}\sum_{i=1}^{N_{2}}\|\hat{y}_{2}^{i} - y_{2}^{i}\|_{2}^{2},
    \label{eq:regression_loss}
\end{equation}

and 
\begin{equation}
    l_{recon} = \frac{1}{N_{1}}\sum_{i=1}^{N_{1}}\|\hat{x_{1}}^{i} - x_{1}^{i}\|_{2}^{2} + \frac{1}{N_{2}}\sum_{i=1}^{N_{2}}\|\hat{x}_{2}^{i} - x_{2}^{i}\|_{2}^{2},
    \label{eq:recons_loss}
\end{equation}

\noindent where $N_{1}$ and $N_{2}$ are the batch sizes in the two datasets.
The total loss is then obtained as:
\begin{equation}
    \textrm{Total Loss} = l_{reg} + l_{recon} + \alpha\times l_{WML}.
    \label{eq:total_loss}
\end{equation}

Over each training iteration, the loss obtained above is backpropagated through both networks and the weights updated to reduce the training error at the next iteration. $\alpha$ is a hyperparameter that controls the influence of the weight mismatch loss on the training of the two networks. If set too big, it forces the networks to look for the same solution, that might not be optimized for each dataset individually. If set too small, it makes the training of the two networks effectively independent of each other. An intermediate value for $\alpha$ results in our networks learning from each other when it is useful for optimization on their own datasets, and adapting only to their respective dataset when knowledge sharing is not useful.

\section{Results and Discussion}
\label{sec:results}
To demonstrate the utility of our workflow for seismic inversion, we use the open source SEAM and Marmousi 2 Datasets. In this section, we briefly describe the salient features of each dataset. 

\subsubsection{Marmousi 2}
\label{subsubsec:marmousi}
The Marmousi 2 model \citep{Martin2002} is an extension to the original Marmousi model for use in AVO (Amplitude Versus Offset) analyses. The original Marmousi model has been used extensively to validate imaging algorithms. However, it contained only a single reservoir. It was extended in a new dataset called Marmousi 2 to be 17km in width and 3.5km in depth. More hydrocarbons were added in structures of different complexities in the dataset. The amount of stratigraphic detail was also increased. The model is accompanied by synthetic seismic data, which has been obtained by convolutional forward modeling the acoustic impedance of the model with a seismic wavelet.

\subsubsection{SEAM}
\label{subsubsec:seam}
The open-source version of the SEAM phase I project \citep{Fehler2010} comes with a 2-D cross-section located at North 23900 in the complete SEAM model. The cross-section is 35km wide in the East-West direction and 15 km deep. It has been sampled at intervals of 20m E-W and 10m in depth. For this cross-section, the dataset provides the p-velocity, $V_{p}$, and the density, $\rho$ in $m/s$ and $kg/m^{3}$ respectively. The SEAM phase I model was simulated with the aim to capture many of the features typical in the Gulf of Mexico. It has a complex salt body and fine stratigraphy with oil and gas reserves. Model properties were derived from fundamental rock properties like Volume of Shale and sand and shale porosities that follow trends characteristic of the Gulf of Mexico, leading to a very realistic model. The dataset also provides a Reverse Time Migrated, stacked seismic data that has been acquired over a subset of the model geometry. As such, the data contains migration and imaging artifacts that may be found in a real-world seismic survey. 

\subsection{Training Setup}
\label{subsec:training_setup}
We obtain the acoustic impedance models for both Marmousi 2 and SEAM datasets by multiplying together their density and p-velocity models. At uniform intervals over the complete length of the models, we sample 12 impedance pseudologs in SEAM and 51 pseudologs in Marmousi 2. For each pseudolog, we also obtain the seismic image patch centered on the well position, as shown in Figure \ref{fig:network}. Each seismic patch is 7 samples wide centered on the well and 701 samples deep, same as the depth of the pseudologs. Both the impedance pseudologs and their corresponding seismic image patches are standardized to have zero mean and unit variance. 

 We set up identical copies of our 2-D Temporal Convolutional Network architecture, one for each dataset. We choose the architecture to have five temporal blocks having 10, 30, 60, 90, and 120 output channels respectively. The kernel size is chosen to have dimensions of nine and three in depth and width respectively. The regression and reconstruction modules are each chosen to be three layer 2-D convolutional neural networks.
 
 We choose a batch size of 16 training samples for Marmousi 2 and 12 for SEAM. Over each training iteration, the losses described in the previous section on network training are computed and back-propagated through both networks. ADAM \citep{kingma2014adam} is chosen to be the optimizer. ADAM adaptively sets the learning rate during the progression of the training. The initial learning rate is set at 0.001. We also choose a weight decay of 0.0001 to constrain the L2 norm of the weights from getting too large. This helps counter overfitting in the network. We run the training for 900 epochs. 
 
 To compare our approach to baseline methods, we carry out two different sets of control experiments. The first group of control experiments aims to compare the performances of deep learning-based seismic inversion both with and without spatial context. Towards this end, we implement the approaches described in \cite{DasCNNInversion}, \cite{motazSemiSupervisedAcoustic}, and \cite{MustafaTCN} respectively. All of these works model only 1-D seismic trace data in a learning-based setup to perform seismic inversion. For an honest comparison with our own proposed work, we implement our 2-D TCN without weight sharing. Each approach is carried out over both SEAM and Marmousi 2 using the same training setup as described in the last paragraph. 
 
 The second group of control experiments is carried out to compare the efficacy of our proposed transfer learning scheme based on weight sharing to other commonly used transfer learning methodologies. In the first of these conventional transfer learning schemes, we pretrain our 2-D TCN on 51 wells from Marmousi for 100 epochs. This network is then finetuned on 12 wells from SEAM for another 900 epochs. In line with the standard finetuning approaches in the literature, we let the deeper parts of the network - the regression and reconstruction modules - train with an initial learning rate of 0.001. In contrast, the feature extraction block is set to have a much smaller learning rate of 0.00025. This is because deep networks are known to learn general, domain-invariant representations in the earlier layers. We therefore do not expect that the feature representations in the earlier layers learnt by the network on Marmousi will change significantly when finetuned on SEAM. In the second conventional transfer learning strategy, we simply combine the training samples from SEAM and Marmousi 2 and train a single 2-D TCN architecure on this dataset. From hereon, we shall refer to this strategy as 'Combined Learning'.
 
 In the next subsection, we describe the results for each control study as observed on SEAM. This is because SEAM is a harder dataset to perform inversion on. In the context of conventional transfer learning, it may therefore be considered the target dataset.

\subsection{Analysis}
\label{subsec:analysis}
\begin{figure*}
    \centering
    \includegraphics[width=\textwidth]{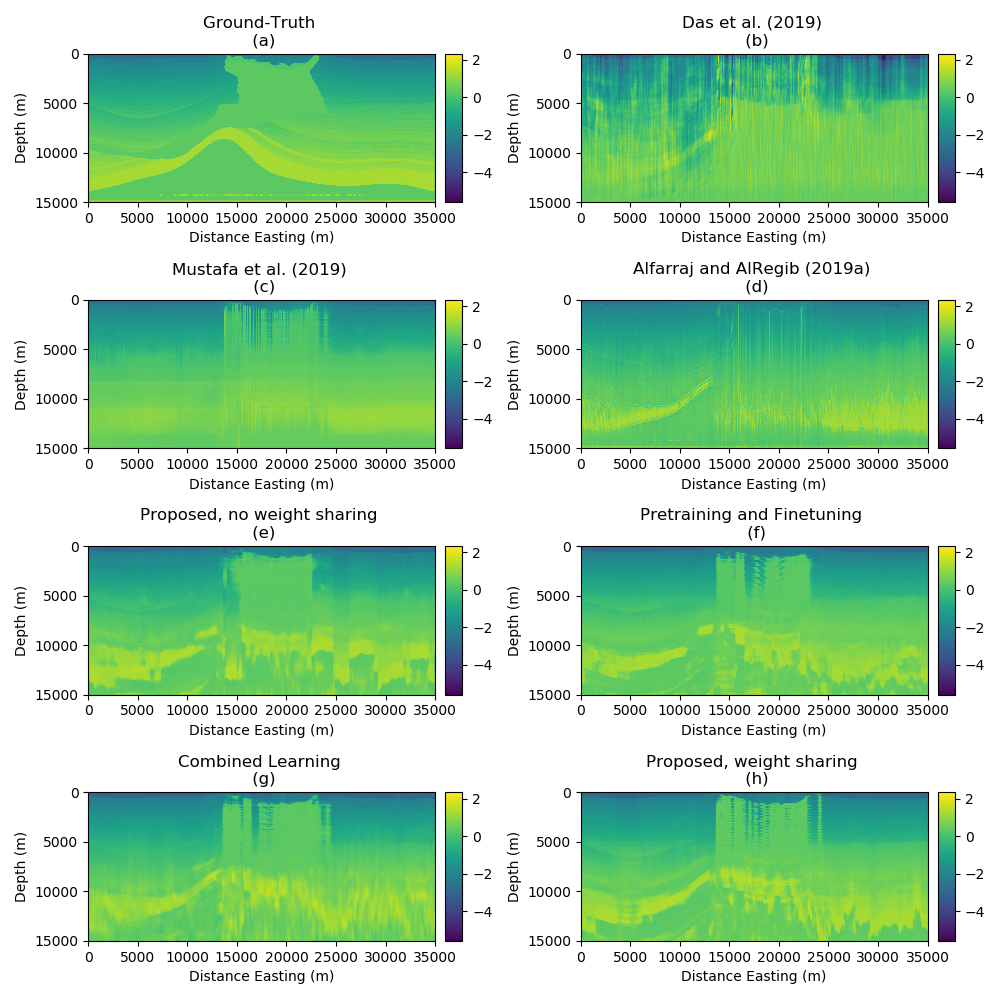}
    \caption{Estimated Acoustic Impedance Profiles for various methods. The ground-truth impedance model is shown at the top-left.}
    \label{fig:seam_plots}
\end{figure*}

We obtain the complete acoustic impedance profiles for all inversion methodologies described in the previous subsection. This is done by evaluating the trained network on all of the seismic data present in the SEAM section. These impedance profiles are shown in Figure \ref{fig:seam_plots}.

One may immediately observe that the plots for 2-D TCN-based approaches (Figure \ref{fig:seam_plots}(e-h)) are better in terms of quality and fidelity to the ground-truth (top left) compared the plots for 1-D models (Figure \ref{fig:seam_plots}(b-d))). The estimations for the 1-D methods are noisier, miss major structures, and suffer with low resolution in terms of stratigraphic details. The next four plots based on 2-D TCNs do much better in terms of all three criteria. This agrees with our intuition that injecting spatial context into the network leads to more robust and laterally consistent estimations - the aim of our first control study.

Looking at the plot by our proposed approach (Figure \ref{fig:seam_plots} h) in closer detail, we observe that it is able to delineate all major structures in the section. Additionally, it has been able to mark out some of the thin stratigraphic variations in the upper left portion of the section, between depths 5000m and 10000m. One can even observe a very faint stratigraphic boundary also present at depth 2500m in the original model. It has also delineated the top of the salt dome reasonably well, considering that we had a very small number of training acoustic impedance pseudologs in this region. It performs well at identifying the boundaries of the high impedance arch present in the lower half of the model. The estimations here are more blurred compared to other regions of the image because of the extremely noisy nature of the seismic data in this region. Not only is the seismic very weak, but sometimes not receptive at all to changes in acoustic impedance. Despite this, one can notice that the transition in the right half around a depth of 5000m has been marked out very close to that in the ground-truth model. Another striking feature is that our estimations are able to preserve a lot of the lateral continuity present in the original model.

\begin{figure*}
  \centering
  \includegraphics[width=\textwidth]{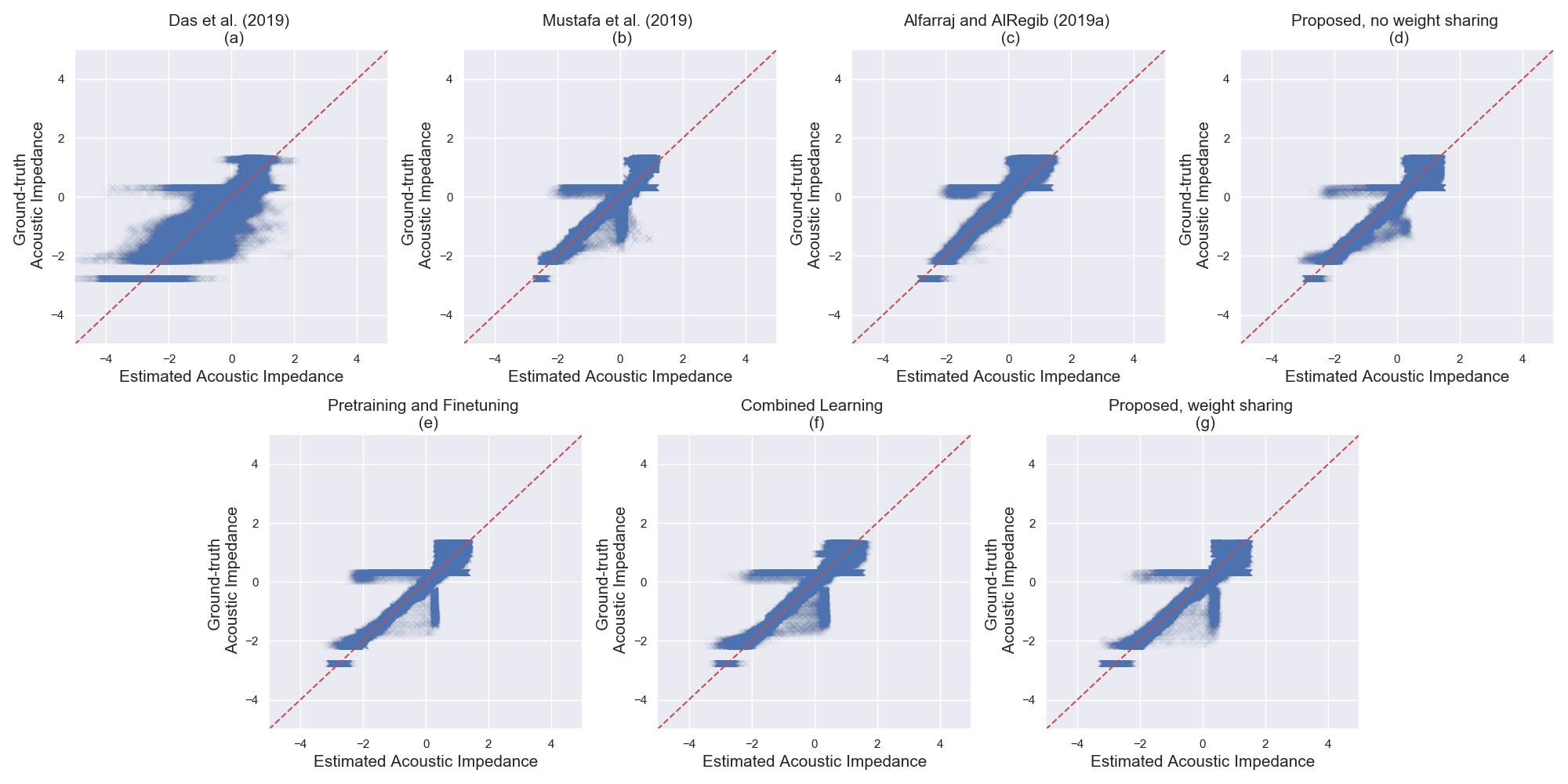}
  \caption{Scatter plots between the Estimated acoustic impedance and the Ground-truth for the different deep learning-based seismic inversion approaches discussed in the paper.}
  \label{fig:scatter}
\end{figure*}

The general picture of superiority of spatial context-based estimations over 1-D modelling is reinforced when one observes the scatter plots in Figure \ref{fig:scatter}. With estimated acoustic impedance on the x-axis and the groundtruth impedance on the y-axis, the scatter plot shows the quality of match between the two quantities. A well-matched plot would have all points lie on the $y=x$ line, shown in red. One may notice how plots by 1-D models (Figure \ref{fig:scatter} (a-c)) tend to be more spread out compared to the 2-D based approaches (Figure \ref{fig:scatter} (d-g)). A case in point is Figure \ref{fig:scatter} (a). Figure \ref{fig:scatter} (b) and (c) tend to do better because being sequence models, they are able to capture the low frequency trend in acoustic impedance along depth for the most part, even if they miss out on many minute details. 

\begin{figure*}
  \centering
  \includegraphics[width=\textwidth]{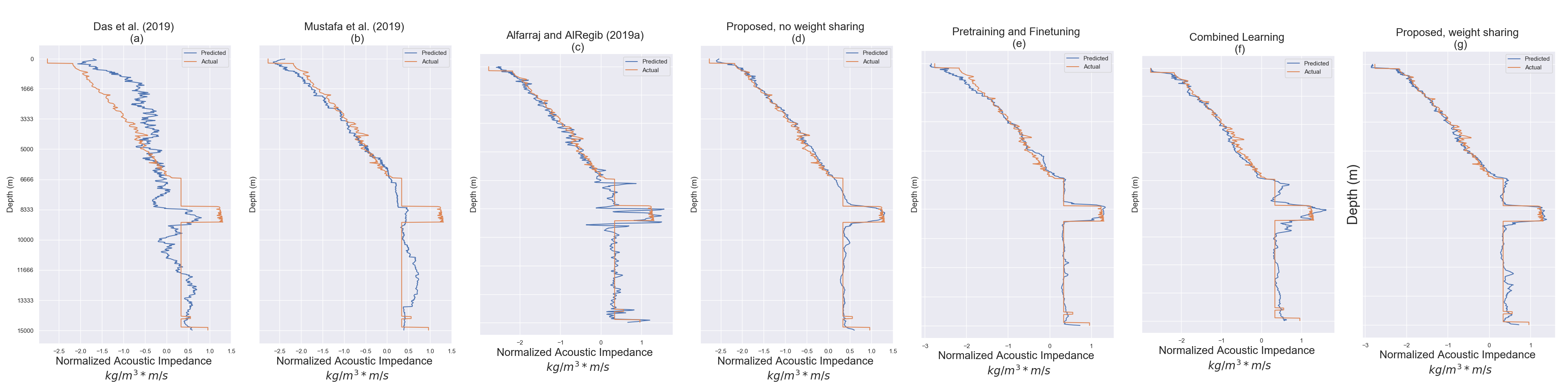}
  \caption{Acoustic Impedance estimations by various methods on the seismic trace present at x=12500m.}
  \label{fig:traces}
\end{figure*}

By taking the impedance pseudolog at $x=12500m$ as a blind well, we compare the the impedance estimations produced by all approaches at this position in Figure \ref{fig:traces}. As expected, the predictions by 2-D TCN match quite closely to the ground-truth trace. Among the 1-D models, \cite{motazSemiSupervisedAcoustic} performs quite well at capturing the variations in this trace, even though the estimations are very oscillatory for the most part. \cite{MustafaTCN} captures the low-frequency trend well, even while it misses many minute details. \cite{DasCNNInversion} performs the least well at capturing both the low frequency trends and the high frequency fluctuations, compared to all other methods.

\begin{figure}
  \centering
  \includegraphics[width=\columnwidth]{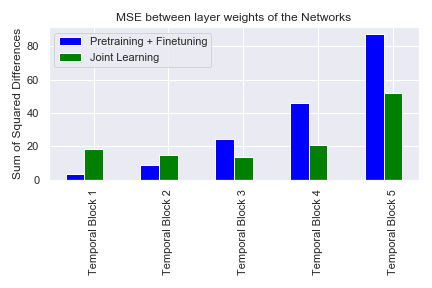}
  \caption{Mean Squared Error between corresponding temporal blocks in the two networks changes along the network depth. Lower error in the first few blocks followed by a higher error in the later ones agrees with our intuition that the networks learn general representations earlier on, followed by more specific features later.}
  \label{fig:weights}
\end{figure}

Before we analyze the results for experiments in the second control study relating to different transfer learning methodologies, it is important to look into the network weights themselves. In our proposed method, we placed a soft constraint on the L2 norm between the corresponding weights in the two networks optimizing on their respective datasets. The idea was to let the networks learn from each other in layers where it is useful to share knowledge and learn dataset specific weights where sharing of knowledge is not useful. We plot the Mean Squared Error between weights in corresponding temporal blocks of the two networks as a bar chart shown in Figure \ref{fig:weights}. It can clearly be observed that the network weights are more similar to each other in the earlier layers before becoming more different in the later ones. This agrees with our intuition of network features changing their nature from general and domain-invariant in earlier layers to dataset- and task-specific in later ones. It is even more remarkable that this behavior was achieved without hand-encoding general and dataset-specific layers in the two networks. As a baseline comparison with the pretraining-based transfer learning, we also plot the differences in the weights of corresponding temporal blocks in our 2-D network before and after finetuning. The trend is much the same, except that we had to hand-encode pre-training layers in the latter case. 

Table \ref{Tab:metrics} shows different accuracy metrics computed over both datasets by the various methods employed. Below, we give a brief definition of these metrics before analyzing the numbers. For all of these, consider we are given a set of estimated measurements, $\{\hat{y}^{i}\}_{i=1}^{N}$, and the corresponding ground-truths, $\{y^{i}\}_{i=1}^{N}$.

\subsubsection{Mean Squared Error (MSE)}
Computes the average mean squared error between the data-points.

\begin{equation}
    MSE = \frac{1}{N}\sum_{i=1}^{N}\|y^{i} - \hat{y}^{i}\|_{2}^{2}
    \label{eq:mse}
\end{equation}

\subsubsection{Mean Absolute Error (MAE)}
Computes the average L1 error between the data-points.
\begin{equation}
    MAE = \frac{1}{N}\sum_{i=1}^{N}|y^{i} - \hat{y}^{i}|
    \label{eq:mae}
\end{equation}
 
\subsubsection{Median Absolute Error (MedAE)}
Computes the median L1 error between the data-points. It is more robust to outliers.
\begin{equation}
    MedAE = Median(|y^{1} - \hat{y}^{1}|,...,|y^{N} - \hat{y}^{N}|)
    \label{eq:medae}
\end{equation}
  
\subsubsection{$r^{2}$ Coefficient}
Measure of goodness of fit. 
\begin{equation}
    r^{2} = 1 - \frac{\sum_{i=1}^{N}(\hat{y}^{i} - y^{i})^{2}}{\sum_{i=1}^{N}(\bar{y}^{i} - y^{i})^{2}}
    \label{eq:r2}
\end{equation}
where $\bar{y}$ is the average of $\{y^{i}\}_{1}^{N}$.

We may now make some comments related to experiments in the second control group. 
For SEAM, our proposed approach performs the best in terms of the MSE and $r^{2}$ coefficient of determination among all methodologies. In terms of MAE and MedAE, it is bested only by the pretraining-based transfer learning scheme, although by a very small margin. It is pertinent to note here that quite a lot of experimentation with the number of pre-training layers and their learning rates had to be done to achieve satisfactory performance with finetuning the pre-trained weights. Even then, we observed large oscillations with the quality of the results from good to poor between successive training runs carried out with the same hyperparameters but different initializations. It may also be observed that network performance after being finetuned on SEAM considerably degrades on Marmousi 2. This serves to reinforce our claim that our method is more robust to different initializations, does not require experimentation with hyperparameters to achieve satisfactory transfer learning performance, and results in increased generalization for all component datasets. 

Moreover, with our proposed approach, one may notice that we achieve a higher generalization performance on \emph{both} datasets (compare the fourth and last entries in the table). As mentioned in the section on weight sharing, the fourth entry corresponds to training the 2-D TCN on each dataset independent of the other dataset. It can be observed that when the weight sharing scheme is incorporated with the 2-D TCN i.e., last entry of the table, we achieve performance gains on both datasets in all metrics. 

One may also observe that the "combined learning" approach is quantitatively worse than the pretraining and the proposed weight sharing-based transfer learning schemes. It is interesting to note that training the 2-D TCN on just the training samples on SEAM (proposed approach with no weight sharing) achieves better quantitative results than training on a combined dataset containg SEAM and Marmousi 2 components (fourth and sixth entries of the table respectively). Curiously, for Marmousi 2, we observe the opposite i.e., including training samples from SEAM into the dataset for Marmousi 2 improves generalization on Marmousi 2. This is because when a single network is trained on training samples coming from different datasets, it is hard to control or predict network behavior. Depending on many factors including network initialization, we may obtain a solution in the weight space that either 1) achieves higher generalization for all component datasets, 2) generalizes to only one component at the expense of another, or 3) less generalizable to all components. In our proposed weight sharing scheme, we are able to significantly increase the likelihood of (1) happening.

One may also observe a curious result for the third row in the Marmousi 2's case: the 1-D sequence modeling approach by \cite{motazSemiSupervisedAcoustic} outperforms all other approaches in all metrics.  This can be explained by noting that the Marmousi 2's synthetic seismic data, as discussed before, has been generated by simple convolutional forward modeling of the acoustic impedance model. \cite{motazSemiSupervisedAcoustic} explicitly simulate the forward model via a CNN. This strongly regularizes the network to perform high quality well log estimations. Notice how even a simple two-layer CNN as used by \cite{DasCNNInversion} produces reasonable estimations on Marmousi 2. Both however, struggle to produce output of a similar quality on SEAM, where the seismic dataset is much noisier. Our own proposed approach, in contrast, comes in a close second.

\begin{table*}
\centering
  \begin{tabular}{m{5cm} m{1cm} m{1cm} m{1cm} m{1cm} m{1cm} m{1cm} m{1cm} m{1cm}} \toprule

    \multirow{2}{*}{Method} &
      \multicolumn{4}{c}{SEAM} &
      \multicolumn{4}{c}{Marmousi 2} \\
    & MSE & MAE & MedAE & $r^{2}$ & MSE & MAE & MedAE & $r^{2}$\\ \midrule

     \cite{DasCNNInversion} & 0.3013 & 0.4097 & 0.3252 & 0.7123 & 0.0652	& 0.1827 &	0.1282	& 0.9345 \\

     \cite{MustafaTCN} & 0.1049 &	0.2122&	0.1411 & 0.8842 & 0.0831 &	0.2156 & 0.1631	& 0.9072\\

     \cite{motazSemiSupervisedAcoustic} &0.1964&	0.2678	&0.1377&	0.7774&	\textit{0.0161}	&\textit{0.0822}&	\textit{0.0531}&	\textit{0.9841}\\

     Proposed, no weight sharing &0.0996&	0.1874&	0.0932&	0.8974&	0.0361&	0.1307&	0.0948&	0.9619\\
     
     Pretraining and Finetuning & 0.1040& 0.1730 & 0.0828 & 0.8983& 0.4931& 0.5805& 0.4870 &0.4912 \\
     
     Combined Learning & 0.1125& 0.2035& 0.1126& 0.8945& 0.0346 &0.1275 & 0.0862& 0.9664 \\
     
     Proposed, weight sharing &\textbf{0.0966} &	0.1781	&0.0914	&\textbf{0.9041}&	\textbf{0.0292}	&\textbf{0.1032}	&\textbf{0.0596}&	\textbf{0.9701} \\ \bottomrule

  \end{tabular}
  \caption{Regression accuracy metrics for the various methods discussed in the paper. We compute and display, for both SEAM and Marmousi 2, the Mean Squared Error (MSE), Mean Absolute Error (MAE), Median Absolute Error (MedAE), and the $r^{2}$ coefficient of determination between the estimated acoustic impedance produced by the method and the ground truth acoustic impedance for the dataset.}
  \label{Tab:metrics}
\end{table*}






     
     
     



\section{Conclusion}
\label{sec:conclusion}
Noise and large geological variations in seismic data, coupled with a limited number of well logs can significantly degrade the performance of deep learning-based seismic inversion algorithms. Most methods in the literature work in a trace-by-trace fashion on seismic data, inverting each trace independently of other traces to obtain rock property traces. This leads to lateral discontinuities in the estimated rock property section. We propose a methodology whereby the local spatial context is injected into the network that also models seismic data temporally for more regularized estimations of rock property. We also demonstrate a scheme whereby mutually beneficial information from multiple datasets can be utilized to improve individual estimations on all datasets. We demonstrate our workflow for acoustic impedance estimation on SEAM and Mamrousi datasets, achieving $r^{2}$ coefficients of 0.9041 and 0.9701 respectively. We compare and contrast our approach to other existing approaches in the literature and show how our method produces more robust estimations in the presence of noisy seismic data and limited well logs. One can use this approach to estimate other rock properties. Moreover, this can be extended to incorporate 3-D spatial context in 3-D seismic volumes. We can also scale this up to learning from more than two datasets. 

\bibliographystyle{seg}  

\end{document}